\documentclass{article}
\usepackage{ijcai25}

\usepackage{times}
\usepackage{soul}
\usepackage{url}
\usepackage[hidelinks]{hyperref}
\usepackage[utf8]{inputenc}
\usepackage[small]{caption}
\usepackage{graphicx}
\usepackage{amsmath}
\usepackage{amsthm}
\usepackage{booktabs}
\usepackage{algorithm}
\usepackage{algorithmic}
\usepackage[switch]{lineno}
\usepackage{multirow}

\usepackage{xcolor}
\usepackage{amssymb}
\usepackage{booktabs}
\usepackage{makecell}
\renewcommand{\arraystretch}{0.8}
\title{\emph{ST-USleepNet}: A Spatial-Temporal Coupling Prominence Network for Multi-Channel Sleep Staging}

\author{
Jingying Ma$^1$
\and
Qika Lin$^1$ \footnote{Corresponding Authors} \and
Ziyu Jia$^{3,4}$ \textsuperscript{*} \And
Mengling Feng$^{1,2}$\\
\affiliations
$^1$Saw Swee Hock School of Public Health, National University of Singapore, Singapore\\
$^2$Institute of Data Science, National University of Singapore, Singapore\\
$^3$Beijing Key Laboratory of Brainnetome and Brain-Computer Interface, Institute of Automation, Chinese Academy of Sciences, Beijing, China\\
$^4$Brainnetome Center, Institute of Automation, Chinese Academy of Sciences, Beijing, China\\
\emails
jingyingma@u.nus.edu,
qikalin@foxmail.com,
jia.ziyu@outlook.com,
mornin@nus.edu.sg
}

\makeatletter
\long\def\@makecaption#1#2{
  \vskip\abovecaptionskip
  \sbox\@tempboxa{\normalfont\small #1. #2}%
  \ifdim \wd\@tempboxa >\hsize
    {\normalfont\small #1. #2\par}
  \else
    \global \@minipagefalse
    \hb@xt@\hsize{\hfil\box\@tempboxa\hfil}
  \fi
  \vskip\belowcaptionskip}
\makeatother

\begin{document}

\maketitle

\begin{abstract}
    Sleep staging is critical to assess sleep quality and diagnose disorders. Despite advancements in artificial intelligence enabling automated sleep staging, significant challenges remain: (1) Simultaneously extracting prominent temporal and spatial sleep features from multi-channel raw signals, including characteristic sleep waveforms and salient spatial brain networks. (2) Capturing the spatial-temporal coupling patterns essential for accurate sleep staging. To address these challenges, we propose a novel framework named ST-USleepNet, comprising a spatial-temporal graph construction module (ST) and a U-shaped sleep network (USleepNet). The ST module converts raw signals into a spatial-temporal graph based on signal similarity, temporal, and spatial relationships to model spatial-temporal coupling patterns. The USleepNet employs a U-shaped structure for both the temporal and spatial streams, mirroring its original use in image segmentation to isolate significant targets. Applied to raw sleep signals and graph data from the ST module, USleepNet effectively segments these inputs, simultaneously extracting prominent temporal and spatial sleep features. Testing on three datasets demonstrates that ST-USleepNet outperforms existing baselines, and model visualizations confirm its efficacy in extracting prominent sleep features and temporal-spatial coupling patterns across various sleep stages. The code is available at https://github.com/Majy-Yuji/ST-USleepNet.
\end{abstract}

\section{Introduction}
Sleep disorders are associated with a wide range of health issues \cite{mahowald2005insights}, including cardiovascular diseases \cite{redline2023obstructive}, metabolic disorders \cite{gileles2016biological}, and neurodegenerative conditions \cite{abbott2016chronic}. Accurate diagnosis and management of these disorders heavily rely on sleep staging \cite{malhotra2002obstructive}, a process that involves categorizing sleep into distinct stages based on physiological signals. This classification provides essential insights into sleep patterns and plays a critical role in identifying abnormalities.

\begin{figure}[h!]
\centering
\includegraphics[width = 1\columnwidth]{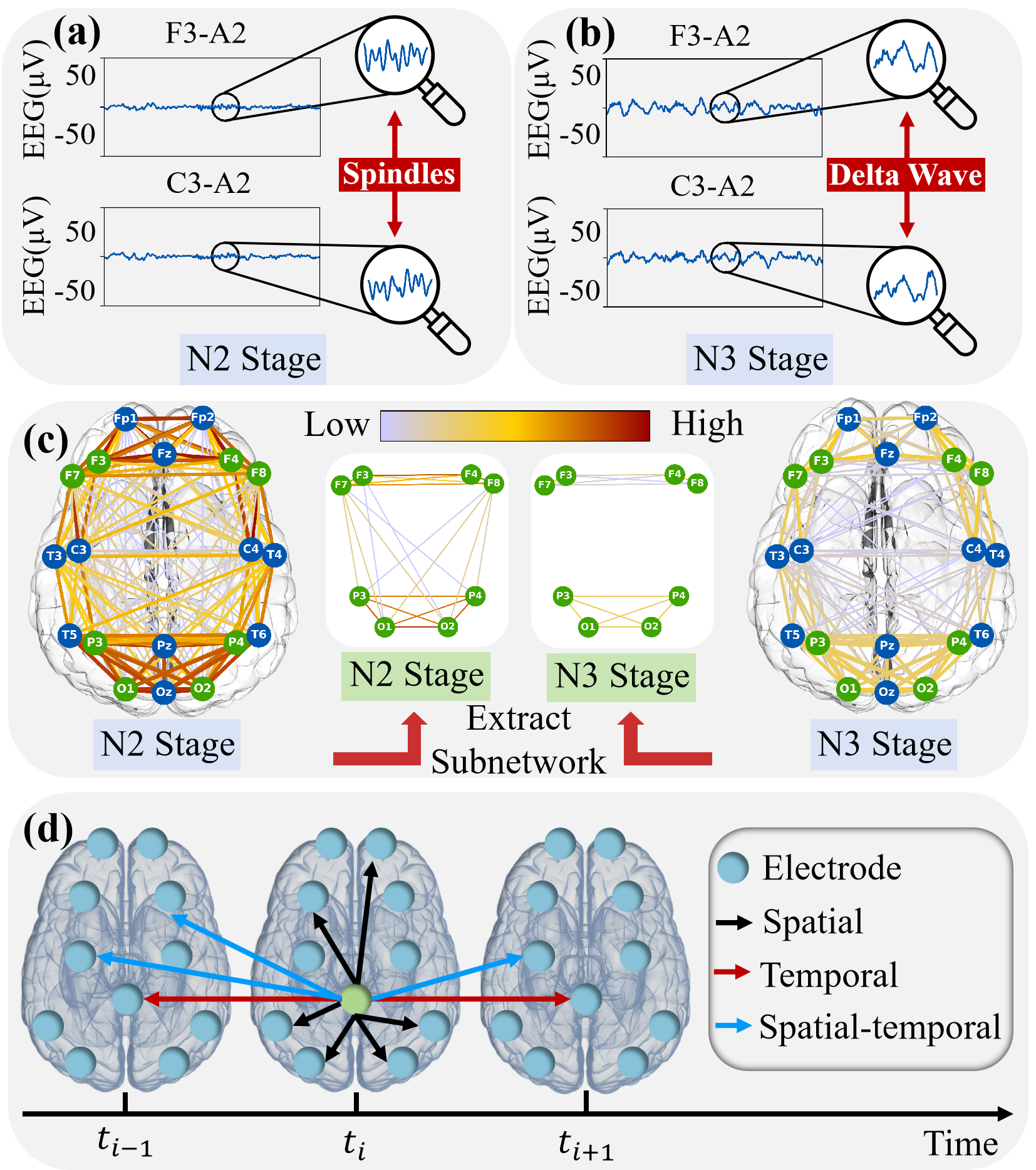}
\caption{Illustration of brain activities across sleep stages: (a–b) display characteristic sleep waveforms from different stages. (c) presents brain networks and salient subnetworks during N2 and N3. (d) illustrates different types of spatial-temporal relationships.}
\label{fig: intro}
\end{figure}

Sleep staging is traditionally performed using polysomnography (PSG), which primarily analyzes electroencephalogram (EEG) signals, but also incorporates other physiological measurements such as electrooculogram (EOG) and electromyogram (EMG). These signals are segmented into 30-second epochs and manually classified by experts into standard stages, including wake, non-REM (N1, N2, N3) sleep stages and rapid eye movement (REM) \cite{berry2012rules}. However, this manual process is labor-intensive and challenging to scale for large datasets \cite{danker2009interrater}. To improve the efficiency and accuracy of sleep staging, researchers have turned to artificial intelligence to develop automated sleep staging models \cite{2017DeepSleepNet,Perslev2019,2020TinySleepNet}. Despite advances, there are two unresolved challenges in this field.

\textit{How to simultaneously capture the characteristic sleep waveforms and salient spatial brain networks from multi-channel raw signals?} Different sleep stages exhibit prominent temporal and spatial features, primarily the characteristic sleep waveforms and salient spatial brain networks as outlined by the AASM sleep standards \cite{berry2012rules}. Figures \ref{fig: intro} (a) and (b) show characteristic sleep waveforms over time. Panel (a) features spindle waves from the N2 stage, while panel (b) displays delta waves from the N3 stage. Figure \ref{fig: intro} (c) presents the brain networks and salient spatial subnetworks during both N2 and N3 stages, with the salient subnetworks highlighted by retaining only the green nodes. Given the crucial information contained in these prominent sleep features, many studies \cite{jia2021b,pei2024wavesleepnet,wang2024subject} have effectively extracted salient waveforms from sleep signals. However, these approaches are limited to single-channel signals, whereas PSG signals are predominantly multi-channel. The prominent sleep features in multi-channel signals are significantly more complex than those in single-channel signals. This complexity has two main sources. One arises from the characteristic sleep waveforms observed across different channels, as illustrated in Figures \ref{fig: intro} (a) and (b), which show simultaneous waveforms in two channels. The other source comes from the formation of distinct functional connections at various sleep stages due to interactions between signals \cite{stevner2019discovery,parente2020functional}. These interactions result in intricate spatial patterns that further enrich the complexity of multi-channel signals. Consequently, simultaneously capturing both multi-channel characteristic sleep waveforms and salient spatial brain networks is crucial for accurate sleep staging and presents a significant challenge.

\textit{How to capture the spatial-temporal coupling patterns to better understand brain activities?} Signals from different brain regions exhibit complex yet crucial spatial-temporal coupling patterns as they evolve over time \cite{Hövel2020}. Figure \ref{fig: intro} (d) illustrates the spatial-temporal relationships within the brain network over time. In the figure, each circle represents a signal channel, corresponding to an electrode placement on the brain; each arrow signifies a connection within the brain network over time. Using the green node as an example, different arrows indicate distinct types of relationships: red arrows denote temporal correlations within the same channel across different time steps, black arrows symbolize spatial dependencies at the same time step across various channels, and blue arrows represent spatial-temporal coupling patterns between different channels at different times. These spatial-temporal connections within the brain are crucial for revealing brain activities \cite{Pang2023}. Most existing models can only capture temporal correlations \cite{2018Automatic,2021An,ji2023,ji2024} or spatial dependencies \cite{jia2021c,ji2022}. Therefore, capturing these spatial-temporal coupling patterns for sleep staging presents a challenge.

To address the above challenges, we propose \emph{ST-USleepNet}, a raw-signal-based novel spatial-temporal coupling prominence framework for sleep staging. It consists of two main components: a spatial-temporal graph construction module (ST) and a U-shaped sleep network (USleepNet). Inspired by studies from other domains \cite{song2020spatial,gong2023astdf}, the ST module converts multi-channel raw signals into a spatial-temporal graph, allowing for the modeling and capturing of spatial-temporal coupling patterns. On the other hand, the USleepNet, which leverages a U-shaped structure originally developed for image segmentation \cite{ronneberger2015u}, segments the signals and the graph constructed by the ST to obtain prominent temporal and spatial sleep features like how image segmentation isolates the main target of an image. Therefore, the USleepNet has two streams: for the temporal stream, we utilize a fully convolutional U-Net to segment characteristic sleep waveforms; for the spatial stream, we employ a graph convolutional U-Net to isolate salient spatial brain networks. The key contributions of this work include the following:

\begin{itemize}
  \item We develop the \emph{USleepNet} module, comprising a temporal prominence network and a spatial prominence network to simultaneously extract multi-channel characteristic sleep waveforms and salient spatial brain networks.
  \item We design a \emph{spatial-temporal graph construction} module to model the spatial-temporal coupling patterns.
  \item Experimental results indicate that our ST-USleepNet model reaches state-of-the-art performance across three sleep staging datasets.
  \item Visualization Studies demonstrate that our method effectively extracts prominent sleep features and spatial-temporal coupling patterns across different sleep stages, providing a certain degree of interpretability.
\end{itemize}

\section{Related Work}

\begin{figure*}[h!]
\centering
\includegraphics[width = 2\columnwidth]{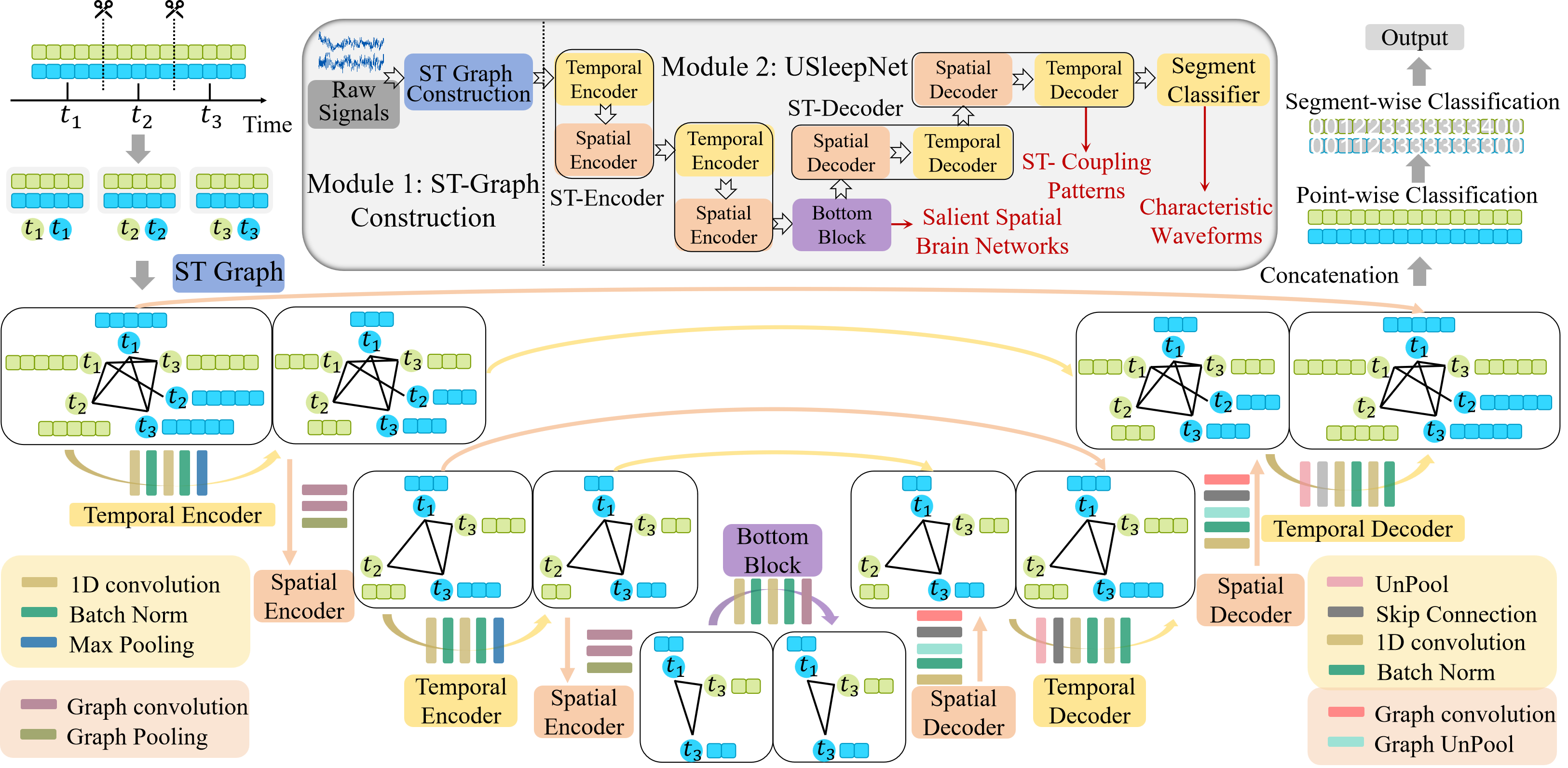}
\caption{ST-USleepNet overview: central box shows the framework; surrounding parts illustrate layers and a two-channel example.}
\label{fig: main}
\end{figure*}

Traditional sleep staging relied on machine learning methods like Random Forest (RF) \cite{Memar2018} and Support Vector Machine (SVM) \cite{Lv2016,Alickovic2018}, which required manually crafted features based on expert knowledge, making the process costly and inefficient. To overcome this, deep learning approaches such as convolutional neural networks (CNNs) \cite{2021An,jia2021a,ji2023,ji2024} and Recurrent Neural Networks (RNNs) \cite{pei2024} were introduced to automatically extract high-dimensional features. However, these methods still follow a two-stage pipeline and lack end-to-end training capabilities.

Recent efforts have focused on end-to-end architectures for processing temporal correlations in sleep staging without preliminary steps. Temporal modeling approaches include CNNs \cite{Perslev2019,jia2021b,perslev2021u}, RNNs \cite{2018Automatic}, their combinations \cite{2017DeepSleepNet,2020TinySleepNet}, and Transformers \cite{Phan2022}. For instance, U-time \cite{Perslev2019} employed a CNN-based Unet for single-channel signals, while SalientSleepNet \cite{jia2021b} integrated a $U^2$-structured CNN Unet with attention to identify typical waveforms. Some studies have also focused on extracting typical waveforms to perform expert-like classification \cite{pei2024wavesleepnet,wang2024subject}. Despite their advancements, these methods primarily focus on single-channel or abstract temporal features, overlooking spatial dependencies.

Spatial dependencies among brain regions are critically important \cite{Sakkalis2011}. GraphSleepNet \cite{jia2021c} was the first to apply a spatial-temporal graph convolution network (ST-GCN) for sleep staging. Subsequently, researchers introduced more graph-based methods \cite{jia2021a,ji2022}. Besides, Liu et al. proposed the BSTT \cite{liu2023} model, which uses a spatial-temporal transformer to capture spatial brain features. However, these spatial methods mainly focus on capturing global spatial features instead of the salient spatial brain networks and ignore the spatial-temporal coupling patterns.

To address the limitations of existing methods and effectively capture characteristic sleep waveforms, salient spatial brain networks, and spatial-temporal coupling patterns across different sleep stages from multi-channel raw signals, we introduce ST-USleepNet for sleep staging.

\section{Methodology}

The framework of ST-USleepNet is illustrated in Figure \ref{fig: main}, consisting of two modules. First, a \emph{spatial-temporal graph construction} module is developed to capture the spatial-temporal coupling patterns, converting raw signals into a spatial-temporal graph. Next, we introduce \emph{USleepNet} to extract characteristic sleep waveforms and identify spatial salient brain networks from these signals. USleepNet is composed of two interwoven streams: the temporal stream, also called the temporal prominence network, represented by yellow blocks in Figure \ref{fig: main}, and the spatial stream, also called the spatial prominence network, depicted by orange blocks. Each stream comprises $l$ encoder and $l$ decoder blocks, with $l=2$ exemplified in the figure. Additionally, the temporal stream integrates a segment classifier. To effectively combine the two streams, a fusion strategy is implemented.

\subsection{Spatial-Temporal Graph Construction}

The spatial-temporal graph construction module aims to transform the raw signals into a spatial-temporal graph. Given a multi-channel physiological signal data $s^{C \times f_sT}$, where $C$ represents the number of channels, $f_s$ denotes the sampling frequency, and $T$ indicates the duration of each segment, the data $s^{C \times f_sT}$ is divided into $n$ patches, resulting in $X^{nC \times f_st}$. We use the raw signals $X$ directly as the initial features of the graph, then we have $G_s = (V, A^{nC \times nC}, X^{nC \times f_st})$, where $V$ represents the set of nodes, $A$ is the adjacency matrix. The nodes of the graph $G_s$ are constituted by patches: node $N_{c_it_j}$ represents a channel $c_i$, where $i \in \{k \in \mathbb{Z} \mid 0 \leq k < C\}$, at a certain time step $t_j$, where $j \in \{k \in \mathbb{Z} \mid 0 \leq k < n\}$. Next, we will discuss how to construct the adjacency matrix $A^{nC \times nC}$.

\begin{figure}[ht]
\centering
\includegraphics[width = 0.9\columnwidth]{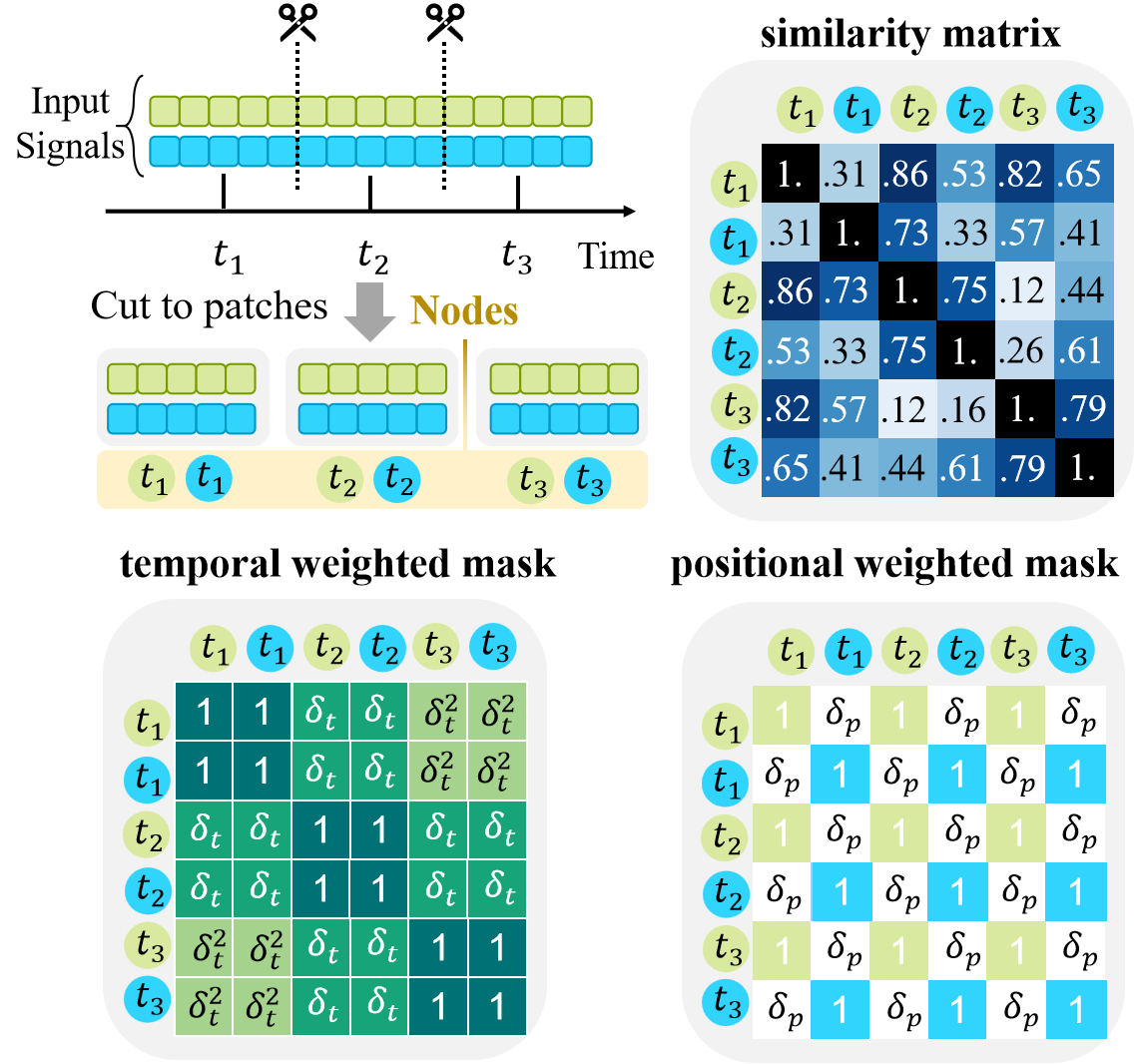}
\caption {Spatial-temporal graph from a 2-channel signal (3 patches/channel); nodes are patches, and adjacency is computed via similarity, temporal, and positional masks.}
\label{fig: st}
\end{figure}

The adjacency matrix $A$ is derived from the multiplication of a similarity matrix with several weighted masks, as shown in Figure \ref{fig: st}. First, an adjacency matrix is established based on the cosine similarity between different features, as highly similar signals often indicate similar functional states. To avoid assigning weights to unrelated nodes, a threshold $\delta$ is applied, establishing edges only between nodes with similarity exceeding $\delta$. The similarity matrix $S$ is then defined as follows:
\begin{equation}
s_{ij} = 
\begin{cases} 
sim\_cos(X_i, X_j) & \text{if } sim\_cos(X_i, X_j) > \delta \text{,} \\
0 & \text{otherwise} \text{.}
\end{cases}
\end{equation}
Here, $X_i$ and $X_j$ represent features of two different nodes; $sim\_cos$ represents the cosine similarity function. Next, we add some prior knowledge using other weighted masks. One is to add the temporal information: patches closer in time should have higher similarity. A temporal weighted mask $Mt$ is constructed based on the time interval between patches. With a hyperparameter $\delta_t \in (0, 1)$, the temporal weight ${mt}_{ij}$ between the feature $X_{c_it_i}$ and $X_{c_jt_j}$ is defined as:
\begin{equation}
{mt}_{ij} = \delta_t^{|t_i - t_j|} \text{,}
\end{equation}
where $t_i$ and $t_j$ are the time steps of the patches, and $t_i,t_j \in \{k \in \mathbb{Z} \mid 0 \leq k < n\}$.
 
Similarly, positional features between channels should be considered because patches from different channels should have lower similarity than those from the same channel. A hyperparameter $\delta_p \in (0, 1)$ is defined to penalize edge weights between different channels, forming a positional weighted mask $Mp$. Thus, the positional weight ${mp}_{ij}$ between the feature $X_{c_it_i}$ and $X_{c_jt_j}$ can be expressed as:
\begin{equation}
{mp}_{ij} = 
\begin{cases} 
\delta_p & \text{if } c_i \neq c_j \text{,} \\
1 & \text{otherwise} \text{.}
\end{cases}
\end{equation}
Therefore, the adjacency matrix $A$ for $G_s$ is:
\begin{equation}
A = S \cdot Mt \cdot Mp \text{.}
\end{equation}

According to the above rules, the multi-channel signal segment $s^{C \times f_sT}$ is transformed into a spatial-temporal graph $G_s = (V, A^{nC \times nC}, X^{nC \times f_st})$.

\subsection{U-shaped Sleep Network}

The spatial-temporal graph $G_s$ constructed from the raw signals is then input into the U-shaped sleep network (USleepNet). There are two interwoven networks, the temporal prominence network for the temporal stream and the spatial prominence network for the spatial stream.

\subsubsection{Temporal Prominence Network}
This network consists of three logical submodules: the temporal encoder block $T_{en}$, the temporal decoder block $T_{de}$, and a segment classifier $T_c$. $T_{en}$ extracts multi-scale temporal features from the feature matrix $X$ of the graph $G_s$, outputting a compressed representation, which $T_{de}$ further processes. Lastly, $T_c$ generates a segment-wise label.

\paragraph{Temporal Encoder Block} $T_{en}$ consists of two 1D convolution layers. The number of channels in the second convolution layer is greater than that in the first layer, since deeper networks tend to capture more complex features. Each convolution uses the same padding, followed by batch normalization. The kernel size remains consistent within the two convolution operations of each $T_{en}$, while it may vary across different $T_{en}$. After these two convolution operations, max pooling is performed to reduce the feature dimensions.

\paragraph{Temporal Decoder Block} $T_{de}$ consists of two transposed-1D convolution layers. Each $T_{de}$ corresponds to a $T_{en}$, performing inverse operations to recover features. It unpools the representation, concatenates it with the corresponding $T_{en}$ output, and applies two sequential convolutions. Kernel sizes are consistent within each $T_{de}$ but may vary across blocks, with batch normalization applied after each convolution.

\paragraph{Segment Classifier} $T_c$ generates segment-wise labels. It takes embedding $H^{ch \times nC \times f_st}$ as input, reshapes it to ${\hat{H}}^{ch \times C \times nf_st}$, and applies a softmax function to produce point-wise labels for each time step and channel. Mean pooling combines the channels into one, and a fully connected layer outputs the segment label.

\subsubsection{Spatial Prominence Network}
For the spatial stream, inspired by Graph U-Net \cite{gao2019graph}, we propose the spatial prominence network. This network follows an encoder-decoder structure and consists of two logical submodules, the spatial encoder block $S_{en}$ and the spatial decoder block $S_{de}$. $S_{en}$ takes the spatial-temporal graph $G_s$ as input and outputs a compressed feature subgraph, while ${S}{de}$ takes the subgraph and recovers a supergraph containing the learned spatial relationships.

\paragraph{Spatial Encoder Block} $S_{en}$ is to extract spatial features between multiple channels and sample important channels to form a new subgraph for high-level feature encoding. It consists of two GCN layers and a graph pooling layer.

The input to $S_{en}$ consists of the adjacency matrix $A^{v \times v}$ and the feature matrix $X^{v \times h}$, where $v$ represents the number of nodes and $h$ represents the feature dimensions. The first graph convolution layer maps $X$ to a higher-dimensional space, and the second graph convolution layer maps $X$ back to its original dimensions. The two consecutive graph convolutions can be represented as follows:
\begin{equation}
\mathbf{X}^{(2)} = Drop(\sigma(\mathbf{A} Drop(\sigma(\mathbf{A} \mathbf{X} \mathbf{W}_1^{h \times \hat{d}})) \mathbf{W}_2^{\hat{d} \times h}) \text{,}
\end{equation}
where $Drop$ means the dropout layer, $W_i$ is a learnable hyperparameter, $d$ is the higher dimension than $h$, and $\sigma$ is ELU, which is defined as:
\begin{equation}
\sigma(x) = \text{ELU}(x) = 
\begin{cases} 
x & \text{if } x > 0 \text{,}\\
\alpha (e^x - 1) & \text{if } x \leq 0 \text{.}
\end{cases}
\end{equation}
Then, a graph pooling layer (gPool) is used to perform downsampling on the graph data. This reduces the size of the graph while retaining information on important nodes. In gPool, we rank the nodes based on importance and select k to generate a subgraph. The computation of gPool can be represented as follows: First, each node's feature vector $X_i$ is projected using a trainable projection vector $P$ to calculate a scalar projection value as an importance score $ic$:
\begin{equation}
ic^v = \frac{X^{v \times h} p^h}{|p^h|} \text{,}
\end{equation}
where $v$ represents the number of nodes, $h$ represents the feature dimension. Then, sort the $ic$ and select k largest values:
\begin{equation}
\text{idx} = \text{rank}(ic, k) \text{,}
\end{equation}
\begin{equation}
\tilde{ic}^k = sigmoid(ic(\text{idx})^k) \text{,}
\end{equation}
where $idx$ is the number of selected top-$k$ nodes, and $\tilde{ic}^k$ is their weights. Feature vectors of the top-$k$ nodes are used to update the adjacency matrix $\tilde{A}$ and feature matrix $\tilde{X}$:
\begin{equation}
\tilde{X}^{k \times h} = X^{v \times h}(\text{idx}, :) \text{,}
\end{equation}
\begin{equation}
\label{eq:update_a}
\tilde{A}^{k \times k} = A^{v \times v}(\text{idx}, \text{idx}) A^{v \times v}(\text{idx}, \text{idx}) \text{,}
\end{equation}
\begin{equation}
\tilde{X}^{k \times h} = \tilde{X}^{k \times h} \odot (\tilde{ic} \mathbf{1}_h^T) \text{,}
\end{equation}
where $\odot$ denotes element-wise multiplication, and $\mathbf{1}_h$ represents an h-dimensional constant vector of 1 used to expand the dimensions of $\tilde{ic}$. In equation \ref{eq:update_a}, the reason for multiplying the two extracted adjacency matrices is to avoid the nodes in the pooled graph becoming isolated as much as possible.

\paragraph{Spatial Decoder Block} $S_{de}$ is used to recover the compressed subgraph obtained by $S_{en}$ into a feature supergraph. It consists of four network layers: a convolution layer, a graph unpooling layer (gUnpool), and two graph convolution layers.

$S_{de}$ takes the adjacency matrix $A$ and the feature matrix $X$ as input. First, a CNN layer is used to extract local information from the feature matrix $X$ and reduce the number of channels in $X$. Next, the data is upsampled using a gUnpool layer. The gUnpool layer corresponds to a gPool layer and uses the node selection information recorded during the downsampling process of the gPool layer. It places the existing nodes and their feature matrices back into their original positions in the graph, while the features of other nodes are set to zero. To avoid significant information loss, a skip connection is used to add the adjacency matrix of the upsampled graph to the corresponding adjacency matrix before downsampling in $S_{de}$. Finally, similar to $S_{en}$, two consecutive graph convolution layers are used: the first maps the feature matrix to a hidden high-dimensional space, and the second restores it to its original dimensions. Through these four layers, the $S_{de}$ block ultimately recovers the compressed feature subgraph into a feature supergraph.

\subsubsection{Spatial-Temporal Block Fusion}
USleepNet comprises an alternating combination of the temporal prominence network and the spatial prominence network. It includes $l$ layers of spatial-temporal encoders (${ST}{en}$), $l$ layers of spatial-temporal decoders (${ST}{de}$), a bottom block (${ST}{b}$), and a segment classifier at the end. Specifically, each ${ST}{en}$ consists of a temporal encoder ($T_{en}$) followed by a spatial encoder ($S_{en}$), while each ${ST}{de}$ consists of a spatial decoder ($S{de}$) followed by a temporal decoder ($T_{de}$). The bottom block (${ST}{b}$) is similar to ${ST}{en}$ but does not include any downsampling. In the encoder phase, it is essential to process with $T_{en}$ first, as PSG data typically has high resolutions and contains noise. Convolution and pooling operations not only reduce dimensions but also remove noise, thereby extracting more representative features. These lower-dimensional features significantly reduce the computational complexity of subsequent graph processing.

\section{Experiments}

\subsection{Datasets}
We evaluate ST-USleepNet on three public datasets. \textbf{ISRUC-S1} \cite{khalighi2016isruc} contains PSG data from 100 patients (55 male and 45 female), with a total of 87,187 sleep segments. The PSG data includes 10 channels of EEG, EOG, and EMG. \textbf{ISRUC-S3} \cite{khalighi2016isruc} contains PSG data from 10 patients (9 male and 1 female), with a total of 8,589 sleep segments. The PSG data includes 10 channels of EEG, EOG, and EMG. \textbf{MASS-SS3} \cite{o2014montreal} contains PSG data from 62 healthy people (28 male and 34 female), with a total of 59,304 sleep segments. The PSG data includes 25 channels of EEG, EOG, EMG, and ECG. For each dataset, the PSG data follows the AASM standards \cite{berry2012rules} and is classified into five categories: Wake, N1, N2, N3, and REM.

\subsection{Experiment Settings}
In this study, we use a cross-subject dataset division and perform 10-fold cross-validation. The evaluation metrics used are accuracy (Acc) and F1-score. All experiments are performed using NVIDIA A100 Tensor Core GPU. We configure the batch size to 64, utilize the Adam optimizer with a learning rate of 1e-3, and train the model for 40 epochs.

\subsection{Baseline Methods}
To comprehensively evaluate the performance of ST-USleepNet, we select 7 models as baselines. These include traditional \textbf{Random Forest (RF)} method \cite{dong2017mixed}; classic CNN-based methods \textbf{DeepSleepNet} \cite{2017DeepSleepNet} and \textbf{TinySleepNet} \cite{2020TinySleepNet}; sleep networks based on U-shaped structures \textbf{U-time} \cite{Perslev2019} and \textbf{SalientSleepNet} \cite{jia2021b}; spatial-temporal sleep network \textbf{GraphSleepNet} \cite{jia2021c}; and the latest general multivariate time series model \textbf{FC-STGCN} \cite{wang2024}.

\subsection{Experiment Results}
The results are shown in Table~\ref{tab:result}, where the best scores are highlighted in bold. Our model achieves state-of-the-art performance across all metrics on all datasets. The largest performance gains are observed on the ISRUC\_S3 dataset. Specifically, ST-USleepNet improves accuracy by +1.14 (1.46\%) compared to the strongest baseline \cite{jia2021c}, and F1-score by +1.02 (1.31\%) compared to the best-performing baseline \cite{jia2021b}. These results underscore the importance of extracting critical features for effective sleep staging.

\begin{table}[ht!]
\centering
\small
\renewcommand{\arraystretch}{1.05}
\begin{tabular}{l|cc|cc|cc}
\hline
\multirow{2}{*}{Model} & \multicolumn{2}{c|}{ISRUC-S1} & \multicolumn{2}{c|}{ISRUC-S3} & \multicolumn{2}{c}{MASS-SS3} \\

\cline{2-7}

 & Acc & F1 & Acc & F1 & Acc & F1 \\

\hline

Random Forest & 71.46 & 70.93 & 72.19 & 71.65 & 80.86 & 80.54 \\

DeepSleepNet & 75.46 & 75.12 & 76.83 & 76.71 & 82.44 & 81.56 \\

TinySleepNet & 75.87 & 75.09 & 77.24 & 76.93 & 82.52 & 82.11 \\

U-time & 74.68 & 74.57 & 74.36 & 73.88 & 82.21 & 81.83 \\

SalientSleepNet & 75.90 & 75.78 & 77.10 & 77.88 & 84.80 & 83.73 \\

GraphSleepNet & 76.35 & 76.67 & 77.97 & 77.73 & 84.86 & 84.27 \\

FC-STGNN & 76.29 & 76.40 & 76.40 & 76.85 & 83.24 & 82.36 \\ 

\hline

\textbf{ST-USleepNet} & \textbf{78.28} & \textbf{77.95} & \textbf{79.11} & \textbf{78.90} & \textbf{85.36} & \textbf{84.88} \\ 

\hline
\end{tabular}
\caption{\normalfont Performance Comparison with Baseline Models.}
\label{tab:result}
\end{table}

\subsection{Ablation Study}
To verify the contribution of the module in ST-USleepNet, we design three variant models by removing the corresponding modules: 1) Without temporal prominence network (-T). 2) Without spatial prominence network (-S). 3) Without spatial-temporal graph construction (-ST). In the `-ST' variant model, we construct a spatial graph using the raw EEG signals in the whole segment based on the cosine similarity.

\begin{table}[ht!]
\centering
\small
\setlength{\tabcolsep}{2pt}
\renewcommand{\arraystretch}{1.05}
\begin{tabular}{c|cc|cc|cc}
\hline
\multirow{2}{*}{Ablation} & \multicolumn{2}{c|}{ISRUC-S1} & \multicolumn{2}{c|}{ISRUC-S3} & \multicolumn{2}{c}{MASS-SS3} \\

\cline{2-7}

 & Acc & F1 & Acc & F1 & Acc & F1 \\

\hline

\textbf{ST-USleepNet} & \textbf{78.28} & \textbf{77.95} & \textbf{79.11} & \textbf{78.90} & \textbf{85.36} & \textbf{84.88} \\ 

\hline

\multirow{2}{*}{-T} & 75.59 & 74.99 & 77.00 & 76.83 & 83.80 & 83.19\\

& -3.44\% & -3.80\% & -2.67\% & -2.62\% & -1.83\% & -1.99\% \\

\hline

\multirow{2}{*}{-S} & 75.77 & 75.41 & 77.40 & 77.05 & 84.07 & 83.59 \\

& -3.21\% & -3.26\% & -2.16\% & -2.34\% & -1.51\% & -1.52\% \\

\hline

\multirow{2}{*}{-ST} & 75.39 & 75.26 & 77.38 & 77.12 & 83.71 & 83.23 \\

& -3.69\% & -2.69\% & -2.19\% & -2.26\% & -1.93\% & -1.94\%\\

\hline
\end{tabular}
\caption{\normalfont Results of Ablation Study.}
\label{tab:alb}
\end{table}

Table \ref{tab:alb} shows the comparison results of the ablation experiments between the variant models and the original model on the three datasets. It can be seen that removing either the temporal prominence network or the spatial prominence network adversely affects the model's performance, and the impact is similar. This indicates that extracting characteristic sleep waveforms and salient spatial brain networks is important. Additionally, removing the temporal-spatial graph construction module also affects the performance, suggesting that we should also focus on the spatial-temporal coupling patterns contained throughout the entire sleep signal, rather than just temporal or spatial features.

\subsection{Hyperparameter Analysis}
We explore the impact of various hyperparameters and identify two with the most significant influence. 

First, ST-USleepNet implements a multi-scale strategy in the temporal prominence network by dynamically varying convolutional kernel sizes and in the spatial prominence network by using different hidden layer dimensions, allowing the network to capture features at multiple scales effectively. To validate that the model adopts the optimal strategy, we experimented with the following combinations: 1)increasing then decreasing (our model's strategy); 2)consistently using small kernels and hidden layers (small); 3)consistently using large kernels (large); 4)decreasing then increasing (reverse). We found that using decreasing convolutional kernels and hidden layer dimensions in the encoder and increasing them in the decoder achieves the best performance, as shown in Table \ref{tab:kernal}.

\begin{table}[ht!]
\centering
\small
\renewcommand{\arraystretch}{1.05}
\begin{tabular}{c|cc|cc|cc}
\hline
\multirow{2}{*}{Strategy} & \multicolumn{2}{c|}{ISRUC-S1} & \multicolumn{2}{c|}{ISRUC-S3} & \multicolumn{2}{c}{MASS-SS3} \\

\cline{2-7}

 & Acc & F1 & Acc & F1 & Acc & F1 \\

\hline

Small & 77.22 & 77.03 & 77.93 & 77.78 & 84.89 & 84.72 \\

Large & 77.92 & 77.56 & 78.64 & 78.02 & 84.74 & 84.09 \\

Reverse & 75.87 & 75.69 & 76.86 & 76.53 & 83.12 & 82.83 \\

\hline

\textbf{ST-USleepNet} & \textbf{78.28} & \textbf{77.95} & \textbf{79.11} & \textbf{78.90} & \textbf{85.36} & \textbf{84.88} \\ 

\hline
\end{tabular}
\caption{\normalfont Comparison across Different Convolution Kernel / Hidden Layer Arrangement Strategies.}
\label{tab:kernal}
\end{table}

Second, we investigate the impact of depth in the U-shaped network on the performance and determine that a four-layer network is the optimal structure, as shown in Table \ref{tab:depth}.

\begin{table}[ht!]
\centering
\small
\renewcommand{\arraystretch}{1.05}
\begin{tabular}{c|cc|cc|cc}
\hline
\multirow{2}{*}{Depth} & \multicolumn{2}{c|}{ISRUC-S1} & \multicolumn{2}{c|}{ISRUC-S3} & \multicolumn{2}{c}{MASS-SS3} \\

\cline{2-7}

 & Acc & F1 & Acc & F1 & Acc & F1 \\

\hline

2 & 74.39 & 74.54 & 75.66 & 75.05 & 82.89 & 82.04 \\

3 & 76.85 & 76.98 & 77.72 & 77.45 & 84.43 & 84.06 \\

4 & 78.28 & \textbf{77.95} & \textbf{79.11} & 78.90 & \textbf{85.36} & \textbf{84.88} \\ 

5 & \textbf{78.65} & 78.12 & 79.07 & \textbf{78.84} & 85.21 & 84.70 \\

\hline
\end{tabular}
\caption{\normalfont Comparison of Different Network Depths.}
\label{tab:depth}
\end{table}

\subsection{Model Visualization}
In ST-USleepNet, each module can extract one kind of sleep feature: the temporal prominence network can capture characteristic sleep waveforms from multiple channels simultaneously, revealing temporal patterns across different sleep stages; the spatial prominence network can capture the salient spatial brain networks, highlighting key brain regions involved in sleep; and the spatial-temporal graph construction module can capture the spatial-temporal coupling patterns, revealing how brain regions interact over time and space. In this section, we show the visualizations for these three types of sleep features, which not only provide insights into the underlying data patterns in different domains but also offer a level of interpretability for the predictive model.

\begin{figure*}[h!]
\centering
\includegraphics[width = 1.9\columnwidth]{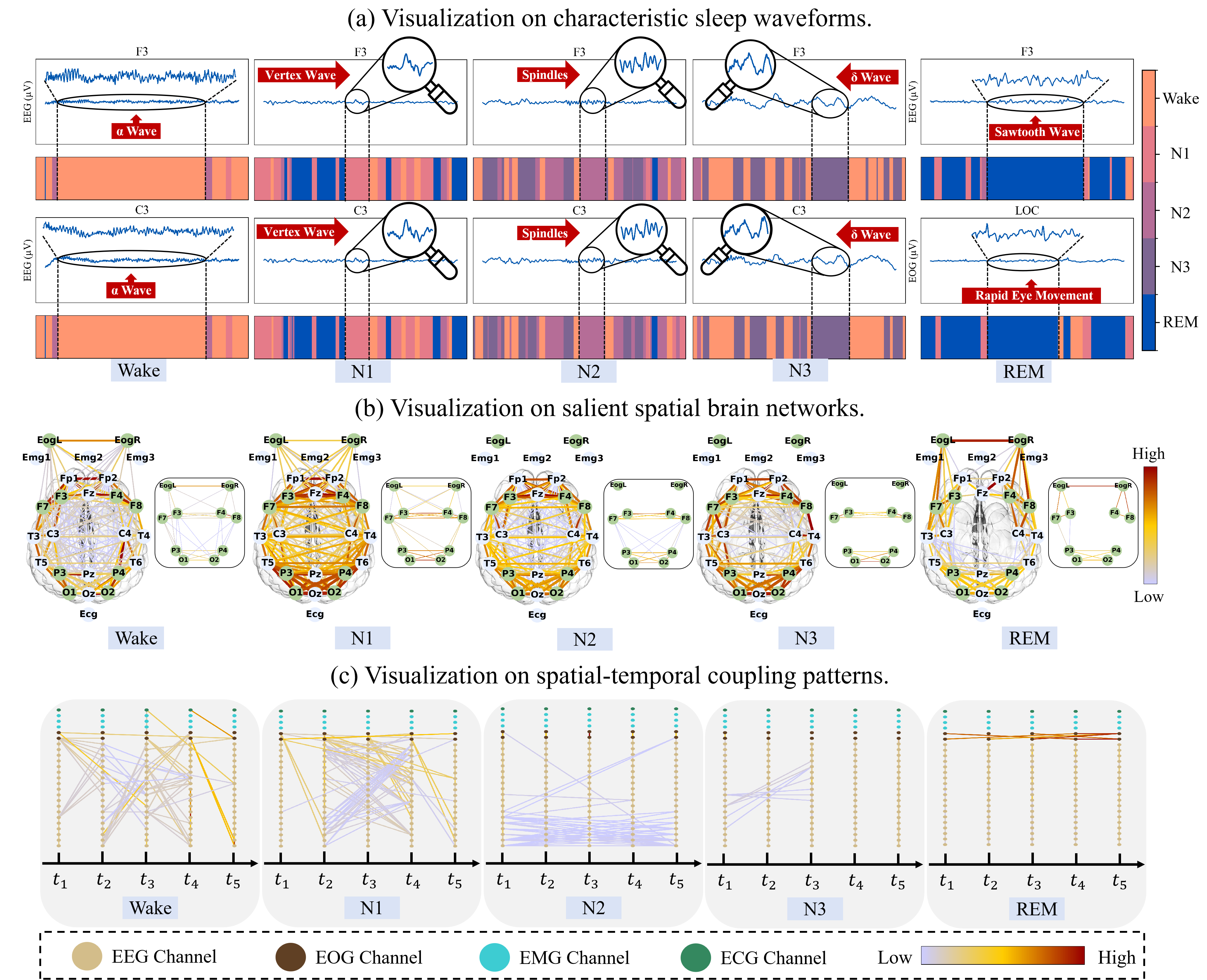}
\caption{Visualizations of sleep features that illustrates three types of sleep features across temporal, spatial, and spatial-temporal domains, highlighting the data patterns extracted by the model. This provides a degree of interpretability for the prediction.}
\label{fig: visual}
\end{figure*}

\subsubsection{Characteristic Sleep Waveforms}
We visualize the model's point-wise labels for each channel, showcasing that ST-USleepNet effectively captures characteristic sleep waveforms across multiple channels for various sleep stages. As shown in Figure \ref{fig: visual} (a), the model identifies typical waveforms for the five sleep stages, with results displayed for two channels per stage. Notably, it captures $\alpha$ waves during the wake period, vertex waves in N1, spindles in N2, $\delta$ waves in N3, and during REM, EEG sawtooth waves and EOG rapid eye movements.

\subsubsection{Salient Spatial Brain Network}
We combine the spatial-temporal graphs generated by the model into a single spatial graph. From the bottom block's output, we select channels with the largest sums of degrees across different time steps within the same channel. Using this information, we extract corresponding subgraphs from the concatenated spatial graph to represent the salient spatial brain network for each sleep stage. The visualization results are shown in Figure \ref{fig: visual} (b). In the figure, the left panel shows the complete brain network, while the right panels display the salient brain networks for each sleep stage. These networks vary significantly: during Wake, brain regions are highly active; in N1, brain activity increases as eye movement decreases; in N2, eye movement further weakens; in N3, brain activity decreases significantly; and during REM, brain activity remains low while eye movement becomes strong.

\subsubsection{Spatial-Temporal Coupling Patterns}
We visualize the spatial-temporal graph output by the model as shown in Figure \ref{fig: visual} (c). In the figure, to highlight the spatial-temporal coupling patterns, the spatial layout at each time step is omitted. Different types of physiological signal channels are represented by distinct colors, and physiological signals from the same time step are arranged in a single column. It is evident that ST-USleepNet effectively captures spatial-temporal coupling patterns, and the patterns reflected are consistent with the salient spatial brain network.

\section{Conclusion}
We propose \emph{ST-USleepNet}, a novel spatial-temporal coupling prominence network for sleep staging. Our model can simultaneously capture characteristic sleep waveforms and salient spatial brain networks across multiple channels. Also, by constructing the raw signals to graph, we capture the spatial-temporal coupling patterns. Experimental results show that ST-USleepNet reaches state-of-the-art performance. Visualization proves that our model has a degree of interpretability.

\section*{Acknowledgments}
This research is supported by A*STAR, CISCO Systems (USA) Pte. Ltd., and the National University of Singapore under its Cisco-NUS Accelerated Digital Economy Corporate Laboratory (Award I21001E0002), the National University of Singapore President's Graduate Fellowship, and the National Natural Science Foundation of China (Grant No. 62306317).

\bibliographystyle{main}
\bibliography{ijcai25}

\begin{thebibliography}{}

\bibitem[\protect\citeauthoryear{Abbott and Videnovic}{2016}]{abbott2016chronic}
Sabra~M Abbott and Aleksandar Videnovic.
\newblock Chronic sleep disturbance and neural injury: links to neurodegenerative disease.
\newblock {\em Nature and science of sleep}, pages 55--61, 2016.

\bibitem[\protect\citeauthoryear{Alickovic and Subasi}{2018}]{Alickovic2018}
Emina Alickovic and Abdulhamit Subasi.
\newblock Ensemble {SVM} method for automatic sleep stage classification.
\newblock {\em {IEEE} Trans. Instrum. Meas.}, 67(6):1258--1265, 2018.

\bibitem[\protect\citeauthoryear{Berry \bgroup \em et al.\egroup }{2012}]{berry2012rules}
Richard~B Berry, Rohit Budhiraja, Daniel~J Gottlieb, David Gozal, Conrad Iber, Vishesh~K Kapur, Carole~L Marcus, Reena Mehra, Sairam Parthasarathy, Stuart~F Quan, et~al.
\newblock Rules for scoring respiratory events in sleep: update of the 2007 aasm manual for the scoring of sleep and associated events: deliberations of the sleep apnea definitions task force of the american academy of sleep medicine.
\newblock {\em Journal of clinical sleep medicine}, 8(5):597--619, 2012.

\bibitem[\protect\citeauthoryear{Danker-Hopfe \bgroup \em et al.\egroup }{2009}]{danker2009interrater}
Heidi Danker-Hopfe, Peter Anderer, Josef Zeitlhofer, Marion Boeck, Hans Dorn, Georg Gruber, Esther Heller, Erna Loretz, Doris Moser, Silvia Parapatics, et~al.
\newblock Interrater reliability for sleep scoring according to the rechtschaffen \& kales and the new aasm standard.
\newblock {\em Journal of sleep research}, 18(1):74--84, 2009.

\bibitem[\protect\citeauthoryear{Dong \bgroup \em et al.\egroup }{2017}]{dong2017mixed}
Hao Dong, Akara Supratak, Wei Pan, Chao Wu, Paul~M Matthews, and Yike Guo.
\newblock Mixed neural network approach for temporal sleep stage classification.
\newblock {\em IEEE Transactions on Neural Systems and Rehabilitation Engineering}, 26(2):324--333, 2017.

\bibitem[\protect\citeauthoryear{Eldele \bgroup \em et al.\egroup }{2021}]{2021An}
Emadeldeen Eldele, Zhenghua Chen, Chengyu Liu, Min Wu, and Cuntai Guan.
\newblock An attention-based deep learning approach for sleep stage classification with single-channel eeg.
\newblock {\em IEEE transactions on neural systems and rehabilitation engineering: a publication of the IEEE Engineering in Medicine and Biology Society}, PP(99), 2021.

\bibitem[\protect\citeauthoryear{Gao and Ji}{2019}]{gao2019graph}
Hongyang Gao and Shuiwang Ji.
\newblock Graph u-nets.
\newblock In {\em international conference on machine learning}, pages 2083--2092. PMLR, 2019.

\bibitem[\protect\citeauthoryear{Gileles-Hillel \bgroup \em et al.\egroup }{2016}]{gileles2016biological}
Alex Gileles-Hillel, Leila Kheirandish-Gozal, and David Gozal.
\newblock Biological plausibility linking sleep apnoea and metabolic dysfunction.
\newblock {\em Nature Reviews Endocrinology}, 12(5):290--298, 2016.

\bibitem[\protect\citeauthoryear{Gong \bgroup \em et al.\egroup }{2023}]{gong2023astdf}
Peiliang Gong, Ziyu Jia, Pengpai Wang, Yueying Zhou, and Daoqiang Zhang.
\newblock Astdf-net: attention-based spatial-temporal dual-stream fusion network for eeg-based emotion recognition.
\newblock In {\em Proceedings of the 31st ACM international conference on multimedia}, pages 883--892, 2023.

\bibitem[\protect\citeauthoryear{Hövel \bgroup \em et al.\egroup }{2020}]{Hövel2020}
Philipp Hövel, Aline Viol, Philipp Loske, Leon Merfort, and Vesna Vuksanovic.
\newblock Synchronization in functional networks of the human brain.
\newblock {\em Journal of Nonlinear Science}, 30, 10 2020.

\bibitem[\protect\citeauthoryear{Ji \bgroup \em et al.\egroup }{2022}]{ji2022}
Xiaopeng Ji, Yan Li, and Peng Wen.
\newblock Jumping knowledge based spatial-temporal graph convolutional networks for automatic sleep stage classification.
\newblock {\em IEEE Transactions on Neural Systems and Rehabilitation Engineering}, 30:1464--1472, 2022.

\bibitem[\protect\citeauthoryear{Ji \bgroup \em et al.\egroup }{2023}]{ji2023}
Xiaopeng Ji, Yan Li, and Peng Wen.
\newblock 3dsleepnet: A multi-channel bio-signal based sleep stages classification method using deep learning.
\newblock {\em IEEE Transactions on Neural Systems and Rehabilitation Engineering}, 31:3513--3523, 2023.

\bibitem[\protect\citeauthoryear{Ji \bgroup \em et al.\egroup }{2024}]{ji2024}
Xiaopeng Ji, Yan Li, Peng Wen, Prabal~Datta Barua, and U.~Rajendra Acharya.
\newblock Mixsleepnet: {A} multi-type convolution combined sleep stage classification model.
\newblock {\em Comput. Methods Programs Biomed.}, 244:107992, 2024.

\bibitem[\protect\citeauthoryear{Jia \bgroup \em et al.\egroup }{2020}]{jia2021c}
Ziyu Jia, Youfang Lin, Jing Wang, Ronghao Zhou, Xiaojun Ning, Yuanlai He, and Yaoshuai Zhao.
\newblock Graphsleepnet: adaptive spatial-temporal graph convolutional networks for sleep stage classification.
\newblock In {\em Proceedings of the Twenty-Ninth International Joint Conference on Artificial Intelligence}, IJCAI, 2020.

\bibitem[\protect\citeauthoryear{Jia \bgroup \em et al.\egroup }{2021a}]{jia2021a}
Ziyu Jia, Youfang Lin, Jing Wang, Xiaojun Ning, Yuanlai He, Ronghao Zhou, Yuhan Zhou, and Li~Wei~H. Lehman.
\newblock Multi-view spatial-temporal graph convolutional networks with domain generalization for sleep stage classification.
\newblock In {\em International Conference of the IEEE Engineering in Medicine and Biology Society}, 2021.

\bibitem[\protect\citeauthoryear{Jia \bgroup \em et al.\egroup }{2021b}]{jia2021b}
Ziyu Jia, Youfang Lin, Jing Wang, Xuehui Wang, Peiyi Xie, and Yingbin Zhang.
\newblock Salientsleepnet: Multimodal salient wave detection network for sleep staging.
\newblock In {\em Proceedings of the Thirtieth International Joint Conference on Artificial Intelligence}, pages 2614--2620. IJCAI, 2021.

\bibitem[\protect\citeauthoryear{Khalighi \bgroup \em et al.\egroup }{2016}]{khalighi2016isruc}
Sirvan Khalighi, Teresa Sousa, Jos{\'e}~Moutinho Santos, and Urbano Nunes.
\newblock Isruc-sleep: A comprehensive public dataset for sleep researchers.
\newblock {\em Computer methods and programs in biomedicine}, 124:180--192, 2016.

\bibitem[\protect\citeauthoryear{Liu and Jia}{2023}]{liu2023}
Yuchen Liu and Ziyu Jia.
\newblock {BSTT:} {A} bayesian spatial-temporal transformer for sleep staging.
\newblock In {\em The Eleventh International Conference on Learning Representations, {ICLR} 2023, Kigali, Rwanda, May 1-5, 2023}. OpenReview.net, 2023.

\bibitem[\protect\citeauthoryear{Lv \bgroup \em et al.\egroup }{2016}]{Lv2016}
Tiantian Lv, Xinzui Wang, Qian Yu, and Yong Yu.
\newblock A features fusion method for sleep stage classification using {EEG} and {EMG}.
\newblock In {\em 4th International Conference on Geo-Informatics in Resource Management and Sustainable Ecosystem, {GRMSE} 2016}, volume 698 of {\em Communications in Computer and Information Science}, pages 176--184. Springer, 2016.

\bibitem[\protect\citeauthoryear{Mahowald and Schenck}{2005}]{mahowald2005insights}
Mark~W Mahowald and Carlos~H Schenck.
\newblock Insights from studying human sleep disorders.
\newblock {\em Nature}, 437(7063):1279--1285, 2005.

\bibitem[\protect\citeauthoryear{Malhotra and White}{2002}]{malhotra2002obstructive}
Atul Malhotra and David~P White.
\newblock Obstructive sleep apnoea.
\newblock {\em The lancet}, 360(9328):237--245, 2002.

\bibitem[\protect\citeauthoryear{Memar and Faradji}{2018}]{Memar2018}
Pejman Memar and Farhad Faradji.
\newblock A novel multi-class eeg-based sleep stage classification system.
\newblock {\em IEEE Transactions on Neural Systems and Rehabilitation Engineering}, 26(1):84--95, 2018.

\bibitem[\protect\citeauthoryear{O'reilly \bgroup \em et al.\egroup }{2014}]{o2014montreal}
Christian O'reilly, Nadia Gosselin, Julie Carrier, and Tore Nielsen.
\newblock Montreal archive of sleep studies: an open-access resource for instrument benchmarking and exploratory research.
\newblock {\em Journal of sleep research}, 23(6):628--635, 2014.

\bibitem[\protect\citeauthoryear{Pang \bgroup \em et al.\egroup }{2023}]{Pang2023}
James Pang, Kevin Aquino, Marianne Oldehinkel, Peter Robinson, Ben Fulcher, Michael Breakspear, and Alex Fornito.
\newblock Geometric constraints on human brain function.
\newblock {\em Nature}, 618:566–574, 5 2023.

\bibitem[\protect\citeauthoryear{Parente and Colosimo}{2020}]{parente2020functional}
Fabrizio Parente and Alfredo Colosimo.
\newblock Functional connections between and within brain subnetworks under resting-state.
\newblock {\em Scientific Reports}, 10(1):3438, 2020.

\bibitem[\protect\citeauthoryear{Pei \bgroup \em et al.\egroup }{2024a}]{pei2024}
Wei Pei, Yan Li, Peng Wen, Fuwen Yang, and Xiaopeng Ji.
\newblock An automatic method using {MFCC} features for sleep stage classification.
\newblock {\em Brain Informatics}, 11(1):6, 2024.

\bibitem[\protect\citeauthoryear{Pei \bgroup \em et al.\egroup }{2024b}]{pei2024wavesleepnet}
Yan Pei, Jiahui Xu, Feng Yu, Lisan Zhang, and Wei Luo.
\newblock Wavesleepnet: An interpretable network for expert-like sleep staging.
\newblock {\em IEEE Journal of Biomedical and Health Informatics}, 2024.

\bibitem[\protect\citeauthoryear{Perslev \bgroup \em et al.\egroup }{2019}]{Perslev2019}
Mathias Perslev, Michael~Hejselbak Jensen, Sune Darkner, Poul~J{\o}rgen Jennum, and Christian Igel.
\newblock U-time: {A} fully convolutional network for time series segmentation applied to sleep staging.
\newblock In {\em Advances in Neural Information Processing Systems 32: Annual Conference on Neural Information Processing Systems 2019, NeurIPS 2019}, pages 4417--4428, 2019.

\bibitem[\protect\citeauthoryear{Perslev \bgroup \em et al.\egroup }{2021}]{perslev2021u}
Mathias Perslev, Sune Darkner, Lykke Kempfner, Miki Nikolic, Poul~J{\o}rgen Jennum, and Christian Igel.
\newblock U-sleep: resilient high-frequency sleep staging.
\newblock {\em NPJ digital medicine}, 4(1):72, 2021.

\bibitem[\protect\citeauthoryear{Phan \bgroup \em et al.\egroup }{2018}]{2018Automatic}
Huy Phan, Fernando Andreotti, Navin Cooray, Oliver~Y. Chén, and Maarten~De Vos.
\newblock Automatic sleep stage classification using single-channel eeg: Learning sequential features with attention-based recurrent neural networks.
\newblock In {\em 40th Annual International Conference of the IEEE Engineering in Medicine and Biology Society (EMBC 2018)}, 2018.

\bibitem[\protect\citeauthoryear{Phan \bgroup \em et al.\egroup }{2022}]{Phan2022}
Huy Phan, Kaare Mikkelsen, Oliver~Y. Chén, Philipp Koch, Alfred Mertins, and Maarten De~Vos.
\newblock Sleeptransformer: Automatic sleep staging with interpretability and uncertainty quantification.
\newblock {\em IEEE Transactions on Biomedical Engineering}, 69(8):2456--2467, 2022.

\bibitem[\protect\citeauthoryear{Redline \bgroup \em et al.\egroup }{2023}]{redline2023obstructive}
Susan Redline, Ali Azarbarzin, and Y{\"u}ksel Peker.
\newblock Obstructive sleep apnoea heterogeneity and cardiovascular disease.
\newblock {\em Nature Reviews Cardiology}, 20(8):560--573, 2023.

\bibitem[\protect\citeauthoryear{Ronneberger \bgroup \em et al.\egroup }{2015}]{ronneberger2015u}
Olaf Ronneberger, Philipp Fischer, and Thomas Brox.
\newblock U-net: Convolutional networks for biomedical image segmentation.
\newblock In {\em Medical image computing and computer-assisted intervention--MICCAI 2015: 18th international conference, Munich, Germany, October 5-9, 2015, proceedings, part III 18}, pages 234--241. Springer, 2015.

\bibitem[\protect\citeauthoryear{Sakkalis}{2011}]{Sakkalis2011}
V.~Sakkalis.
\newblock Review of advanced techniques for the estimation of brain connectivity measured with eeg/meg.
\newblock {\em Computers in Biology and Medicine}, 41(12):1110--1117, 2011.
\newblock Special Issue on Techniques for Measuring Brain Connectivity.

\bibitem[\protect\citeauthoryear{Song \bgroup \em et al.\egroup }{2020}]{song2020spatial}
Chao Song, Youfang Lin, Shengnan Guo, and Huaiyu Wan.
\newblock Spatial-temporal synchronous graph convolutional networks: A new framework for spatial-temporal network data forecasting.
\newblock In {\em Proceedings of the AAAI conference on artificial intelligence}, volume~34, pages 914--921, 2020.

\bibitem[\protect\citeauthoryear{Stevner \bgroup \em et al.\egroup }{2019}]{stevner2019discovery}
ABA Stevner, Diego Vidaurre, Joana Cabral, Kristina Rapuano, S{\o}ren F{\o}ns~Vind Nielsen, Enzo Tagliazucchi, Helmut Laufs, Peter Vuust, Gustavo Deco, Mark~W Woolrich, et~al.
\newblock Discovery of key whole-brain transitions and dynamics during human wakefulness and non-rem sleep.
\newblock {\em Nature communications}, 10(1):1035, 2019.

\bibitem[\protect\citeauthoryear{Supratak and Guo}{2020}]{2020TinySleepNet}
Akara Supratak and Yike Guo.
\newblock Tinysleepnet: An efficient deep learning model for sleep stage scoring based on raw single-channel eeg.
\newblock {\em IEEE}, 2020.

\bibitem[\protect\citeauthoryear{Supratak \bgroup \em et al.\egroup }{2017}]{2017DeepSleepNet}
Akara Supratak, Hao Dong, Chao Wu, and Yike Guo.
\newblock Deepsleepnet: a model for automatic sleep stage scoring based on raw single-channel eeg.
\newblock {\em IEEE Transactions on Neural Systems and Rehabilitation Engineering}, PP(99), 2017.

\bibitem[\protect\citeauthoryear{Wang \bgroup \em et al.\egroup }{2024a}]{wang2024subject}
Jing Wang, Xuehui Wang, Xiaojun Ning, Youfang Lin, Huy Phan, and Ziyu Jia.
\newblock Subject-adaptation salient wave detection network for multimodal sleep stage classification.
\newblock {\em IEEE Journal of Biomedical and Health Informatics}, 2024.

\bibitem[\protect\citeauthoryear{Wang \bgroup \em et al.\egroup }{2024b}]{wang2024}
Yucheng Wang, Yuecong Xu, Jianfei Yang, Min Wu, Xiaoli Li, Lihua Xie, and Zhenghua Chen.
\newblock Fully-connected spatial-temporal graph for multivariate time-series data.
\newblock In {\em Proceedings of the 38th AAAI Conference on Artificial Intelligence}, volume~38, pages 15715--15724, Mar. 2024.

\end{thebibliography}

\end{document}